# Optical image amplification in dual-comb microscopy by use of optical-fiber-amplified interferogram

TAKAHIKO MIZUNO,[1-3] TAKUYA TSUDA,[4] EIJI HASE,[1,3] YU TOKIZANE,[1] RYO OE,[4] HIDENORI KORESAWA,[4] HIROTSUGU YAMAMOTO,[2,3,5] TAKEO MINAMIKAWA,[1-3] AND TAKESHI YASUI[1-3]

[1]*Institute of Post-LED Photonics (pLED), Tokushima University, 2-1 Minami-Josanjima, Tokushima, Tokushima 770-8506, Japan*

[2]*Graduate School of Technology, Industrial and Social Sciences, Tokushima University, 2-1 Minami-Josanjima, Tokushima, Tokushima 770-8506, Japan*

[3]*JST, ERATO, MINOSHIMA Intelligent Optical Synthesizer Project, 2-1 Minami-Josanjima, Tokushima, Tokushima 770-8506, Japan*

[4]*Graduate School of Advanced Technology and Science, Tokushima University, 2-1 Minami-Josanjima, Tokushima, Tokushima 770-8506, Japan*

[5]*Center for Optical Research and Education, Utsunomiya University, 7-1-2, Yoto, Utsunomiya, Tochigi 321-8585, Japan*

**Abstract**

Dual-comb microscopy (DCM), based on a combination of dual-comb spectroscopy (DCS) with two-dimensional spectral encoding (2D-SE), is a promising method for scan-less confocal laser microscopy giving an amplitude and phase image contrast with the confocality. However, signal loss in a 2D-SE optical system hampers increase in image acquisition rate due to decreased signa-to-noise ratio. In this article, we demonstrated optical image amplification in DCM with an erbium-doped fiber amplifier (EDFA). Combined use of image-encoded DCS interferogram and EDFA benefits from not only the batch amplification of amplitude and phase images but also significant rejection of amplified spontaneous emission (ASE) background. Effectiveness of the optical-image-amplified DCM is highlighted in the single-shot quantitative nanometer-order surface topography and the real-time movie of polystyrene beads dynamics under water convection. The proposed method will be a powerful tool for real-time observation of surface topography and fast dynamic phenomena.

## 1. Introduction

An optical frequency comb (OFC) [1-3] has a unique optical spectrum composed of a vast number of discrete, regularly spaced optical frequency modes, and the optical frequency and phase of all OFC modes are secured to a frequency standard by active laser control of carrier-envelope-offset frequency $f_{ceo}$ and a frequency spacing or repetition frequency $f_{rep}$. Dual-comb spectroscopy (DCS) [4-7] has appeared as a new mode to make full use of OFC as an optical frequency ruler for broadband high-precision spectroscopy. Use of two OFCs with slightly different frequency spacings (signal OFC, $f_{rep1}$; local OFC, $f_{rep2} = f_{rep1} + \Delta f_{rep}$) enables us to make a replica of the signal OFC in radio-frequency (RF) region based on a frequency scale of $1:(f_{rep1}/\Delta f_{rep})$, typically $1:10^5$. The resulting mode-resolved OFC spectra of amplitude and phase have been used for broadband high-precision spectroscopy of gas [8,9], solid [10], and thin film [11]. Also, such DCS is available in the broad spectral range of ultraviolet [12], visible [13], mid-infrared [14,15], and terahertz [6,16], due to wavelength diversity of OFC and DCS.

Recently, a new door of application has opened for DCS: dual-comb imaging (DCI) [17-23]. In this case, OFC is regarded as an optical carrier of amplitude and phase with a vast number of discrete frequency channels in place of optical frequency ruler. Then, image pixels to be measured is spectrally encoded into OFC modes by space-to-wavelength conversion or spectral encoding (SE). Finally, image is decoded all at once from the mode-resolved spectrum of the image-encoded OFC acquired by DCS, based on one-to-one correspondence between images pixels and OFC modes. Due to the scan-less imaging capability in DCI and the simultaneous acquisition capability of amplitude and phase spectra in DCS, combination of DCI with confocal laser microscopy enables the scan-less confocal one-dimensional (1D) [17,19-21,23] or two-dimensional (2D) [18,22] imaging of amplitude and phase. Such dual-comb microscopy (DCM) has been effectively applied for the surface topography of a nanometer-scale step-structured sample and the non-staining imaging of standing culture fixed cells [18].

DCM has a potential to boost the image acquisition rate up to $\Delta f_{rep}$ (typically, a few kHz). While such kHz imaging rate will expand application fields of DCM into observation of fast dynamic phenomena, the largely reduced time of image acquisition leads to poor signal-to-noise ratio (SNR). Also, signal loss in a 2D-SE optical system is another reason that hampers increase in the image acquisition rate. Although increase of incident optical power is a straightforward way from the viewpoint of light source, it often causes photodamage in a sample. On the other hand, from the viewpoint of detector, acquisition of DCS interferogram under strong non-interferometric background light makes it difficult to use a highly sensitive photodetector. One interesting approach to enhance SNR in rapid imaging is optical amplification of image-encoded optical signal. For example, the fiber-amplifier-based image amplification was effectively applied for real-time observation of fast dynamic phenomena in serial time-encoded amplified microscopy (STEAM) [24,25].

In this article, we adopted the optical image amplification for DCM to enhance imaging performance in rapid data acquisition or weak signal acquisition. The SNR and contrast in confocal amplitude and phase images were significantly enhanced without influence of incoherent background light of amplified spontaneous emission (ASE) by coherent amplification of image-encoded OFC interferogram in erbium-doped fiber amplifier (EDFA).

## 2. Experimental setup

Figure 1 shows an experimental setup of the optical-image-amplified DCM system. Since the detail of DCM without the optical image amplification is described in the previous paper [18], we here give a brief description of it. We used a pair of homemade femtosecond Er-fiber OFC lasers for a signal OFC (center wavelength = 1560 nm, spectral range = 1545~1575 nm, mean output power = 125 mW, $f_{ceo1}$ = 21.4 MHz, $f_{rep1}$ = 100,388,730 Hz) and a local OFC (center wavelength = 1560 nm, spectral range = 1545~1575 nm, mean output power = 15 mW, $f_{ceo2}$ = 21.4 MHz, $f_{rep2}$ = 100,389,709 Hz, $\Delta f_{rep} = f_{rep2} - f_{rep1}$ = 979 Hz) in DCM. $f_{ceo1}$, $f_{rep1}$, $f_{ceo2}$, and $f_{rep2}$ were all phase-locked to a rubidium frequency standard (not shown, Stanford Research Systems, Inc., Sunnyvale, CA, USA, FS725, frequency = 10 MHz, accuracy = $5 \times 10^{-11}$, instability = $2 \times 10^{-11}$ at 1 s) via a laser control system. The signal OFC beam (mean power = 65 mW) was separated into a reference arm for a reference signal OFC beam and a signal arm for an imaged-encoded signal OFC beam by a 50:50 beam splitter (BS), respectively. The reference signal OFC beam in the reference arm was reflected by a gold plane mirror and was combined with the imaged-encoded signal OFC beam by BS. To separate an interferogram of the reference signal OFC from that of the imaged-encoded signal OFC temporally, we adjusted difference of optical path length between the reference arm and the signal one. In the signal arm, the signal OFC beam was fed into a 2D-SE optical system, composed of a virtually imaged phased array (VIPA, Light Machinery, Inc., Nepean, Ontario, Canada, OP-6721-6743-8, free spectral range = 15.1 GHz, finesse = 110), a diffraction grating (Spectrogon AB, Täby, Sweden, PC 1200 30 × 30 × 6, groove density = 1200 grooves/mm, efficiency = 90 %), and a lens (L1, focal length = 150 mm). The 2D-SE optical system forms 2D spectrograph of signal OFC modes at an optical Fourier plane. The 2D spectrograph was relayed and focused as 2D focal spot array of signal OFC modes onto a sample by a combination of a lens (L2, focal length = 150 mm) with a dry-type objective lens (OL, Nikon Corp., Tokyo, Japan, Plan Apo Lambda 40XC, numerical aperture = 0.95, working distance = 160~250 μm). Reflection, absorption, scattering, and/or phase change of the signal OFC beam in the sample encode the image contrast onto the amplitude and phase spectra of 2D spectrograph. As the image-encoded 2D spectrograph of the signal OFC passed through the same optical system in the opposite direction, each wavelength component of the spectrograph was spatially overlapped with each other again as an image-encoded signal OFC. Typical power of the image-encoded signal OFC was decreased down to several tens to a few hundreds µW mainly due to VIPA passage in the return path. After being spatially overlapped with the reference signal OFC beam by BS, the image-encoded signal OFC beam was fed into the experimental setup of the DCS.

The image-encoded signal OFC beam was spatially overlapped with the local OFC beam in a single-mode fiber coupler (FC). The split ratio of 95:5 in FC was selected due to non-negligible signal loss in the 2D-SE optical system. A polarization controller (PC) in the local OFC path was adjusted to enhance the interferogram signal between the signal and local OFC beams via high polarization overlapping between them in FC. The resulting interferogram signal was amplified by a forward-pumping EDFA composed of a 4.5 m length of erbium-doped fiber (nLIGHT Inc., Vancouver, WA, USA, LIEKKI ER30-4/125, peak core absorption at 1530 nm = 30 dB/m) and a pumping laser diode (Thorlabs Inc., Newton, NJ, USA, BL976-PAG900, wavelength = 980 nm, power = 900 mW). The optically amplified interferogram was detected by a fast photodetector (PD, Thorlabs Inc., Newton, NJ, USA, DET01CFC-N, wavelength = 800~1700 nm, bandwidth = DC to 1.2 GHz). The detected electrical signal was acquired using a digitizer (not shown, National Instruments Corp., Austin, TX, USA, NI PXIe-5122, resolution = 14 bit). The sampling clock signal was synchronized with $f_{rep2}$. The acquired signal corresponded to an interferogram signal with a time widow of 9.96 ns (= $1/f_{rep1}$) and a sampling interval of 97 fs (= $1/f_{rep1}$-$1/f_{rep2}$).

A Fourier transform of the acquired interferogram signal gives the mode-resolved amplitude and phase spectra of the image-encoded signal OFC with a frequency sampling interval of $f_{rep1}$ (=

100,388,730 Hz). Before decoding the confocal amplitude image, a normalized amplitude spectrum was obtained by using the amplitude spectrum of no-image-encoded signal OFC as a reference to eliminate the influence of the spectral shape of the signal OFC. Similarly, we subtracted the phase spectrum of the no-image-encoded signal OFC from the phase spectrum of the image-encoded signal OFC to eliminate the influence of the initial phase in the signal OFC. Then, each data plot of the normalized mode-resolved amplitude and phase spectra was spatially mapped for the confocal amplitude and phase images based on one-to-one correspondence between 2D image pixels and OFC mode [18].

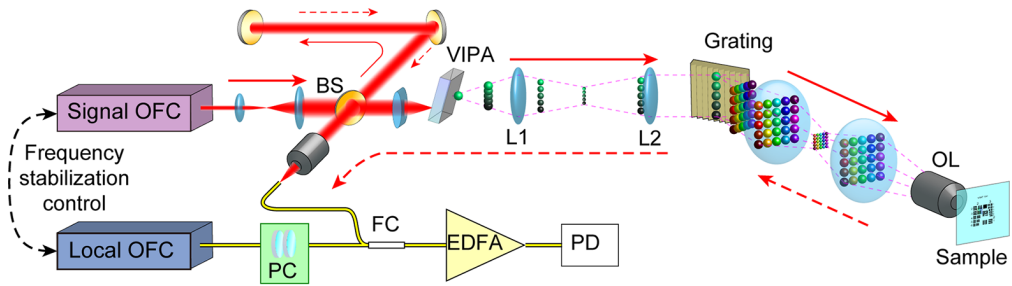

Fig. 1. Experimental setup. BS: beam splitter, VIPA: virtually imaged phased array, L1, L2: lenses (focal length = 150 mm), OL: objective lens, FC: fiber coupler, PC: polarization controller, EDFA: erbium-doped fiber amplifier, PD: fast photodetector.

## 3. Results

*3.1 Basic performance of EDFA for optical image amplification*

We first evaluated the static performance of EDFA using a continuous-wave extended cavity laser diode (ECLD) (Redfern Integrated Optics, Inc., Santa Clara, CA, USA, RIO PLANEX, center wavelength = 1650 nm, FWHM < 2.0 kHz). Figure 2(a) shows a relationship between an optical input power and an optical output power at 1650 nm when a LD pumping current of EDFA was set to 280 mA. A liner relationship was confirmed between them without the saturation, and an optical amplification factor of 13 was achieved from the slope coefficient between them. We used this amplification factor in the following experiments by considering sublinear increase of amplified light, linear increase of ASE, and saturation of the photodetector. Figure 2(b) shows the optical amplification factor with respect to wavelength within the range of OFC spectrum. Such no wavelength dependence will lead to flat amplification of optical image because OFC modes has a one-to-one correspondence with 2D image pixels.

We next evaluate the time-domain performance of EDFA using an interferogram between the signal OFC and the local one. A plane gold mirror was used for a sample. Figures 2(c) and 2(d) compare temporal waveforms of interferogram without and with optical image amplification (time window size = $1/f_{rep1}$ = 9.96 ns, number of signal accumulation = 1,000). Two signals appeared at 1.8 ns and 8 ns in the temporal waveform: the first signal of a single interferogram and the second signal of multiple interferograms. The former was caused by a single reflection on a mirror surface in the reference arm. The later results from multiple reflections in the VIPA in the signal arm, and the time separation between multiple interferograms was equal to an inverse of FSR. From comparison between Figs. 2(c) and 2(d), amplitude of both interferograms was amplified by 13, which is in reasonable agreement with the results in Figs. 2(a) and 2(b).

When the EDFA is used with high gain, ASE is often generated in addition to the amplified input light. Such ASE might be an incoherent background light in the optical image amplification of DCM, leading to degradation of image quality. To evaluate this influence, we evaluate the frequency-domain performance of EDFA. A 1951 USAF resolution test chart with a negative pattern (Edmond Optics, Barrington, NJ, USA, #38-256, spatial frequency: 1.00 lp/mm ~ 228 lp/mm) was used for a sample. Figure 2(e) compares optical spectra of EDFA output with and without the optical input of image-encoded signal OFC, measured by an optical spectrum analyzer (OSA, Yokogawa, AQ6370D-12-D/FC/RFC, wavelength = 600~1700 nm, resolution = 0.02 nm). Fine structure of red plot in Fig. 2(e) reflects the 2D image information of the test chart although the spectral resolution in OSA is insufficient to resolve the detailed spectral features of the image-encoded signal OFC. On the other hand, blue plot of Fig. 2(e) is corresponding to the optical spectrum of ASE. In other words, the optically-amplified image-encoded signal OFC was spectrally overlapped with ASE background in the incoherent measurement of optical spectrum by OSA. However, such ASE background can be significantly rejected by interferometry-based coherent measurement in DCS because ASE is incoherent and is not interfered with the signal OFC or local OFC. Figure 2(f) shows the optical spectrum obtained by Fourier transform of the amplified image-encoded interferogram. The swell of the ASE background was effectively suppressed, and only fine structure of the optical spectrum reflecting the 2D image information was obtained with high contrast. Such ASE rejection capability enables the optical image amplification in DCM without distortion or decease of the image contrast.

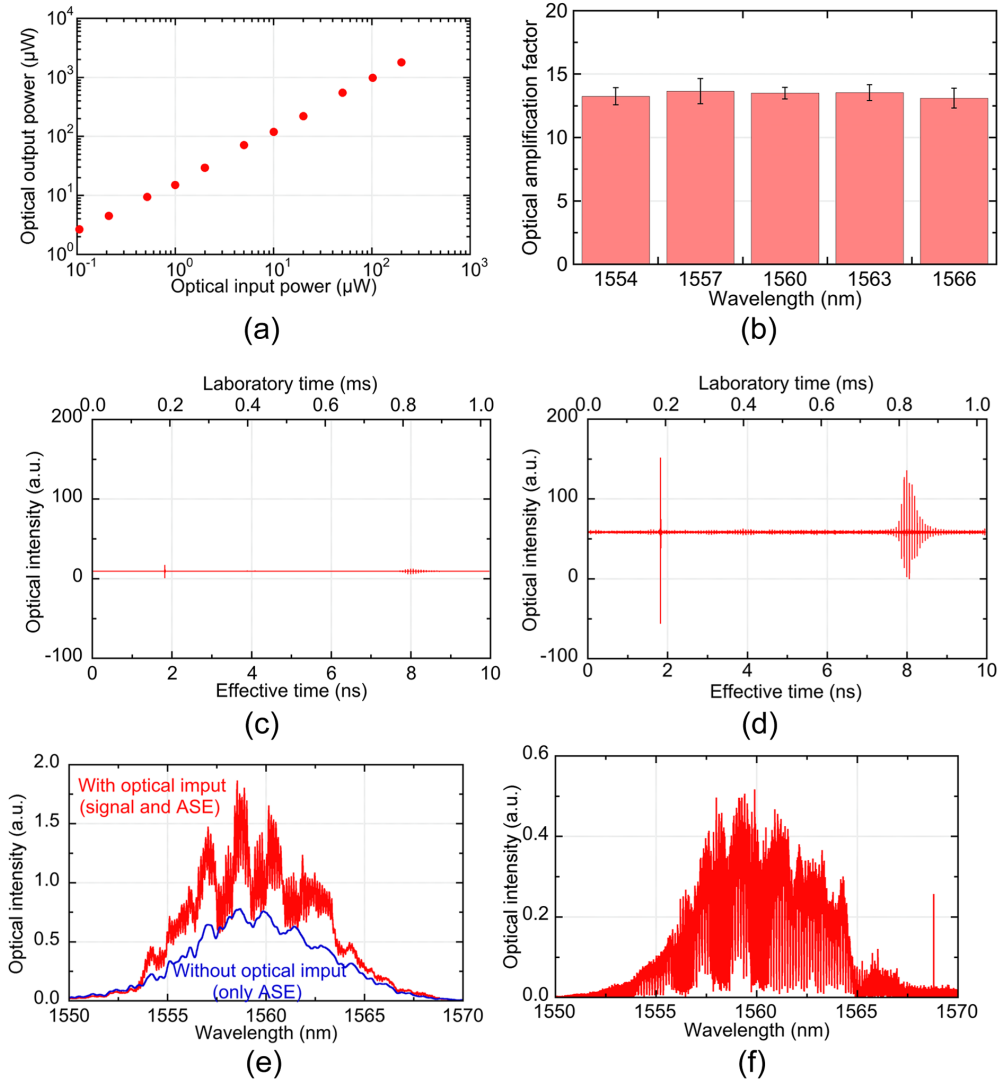

Fig. 2. Basic performance of EDFA. (a) Relationship between an input power and an output power at 1650 nm (pump LD current = 280 mA). (b) Wavelength dependence of optical amplification factor in EDFA. Temporal waveform of interferogram (c) without and (d) with optical image amplification. (e) Comparison of optical spectra of EDFA output with and without the optical input of image-encoded signal OFC measured by an optical spectrum analyzer. (f) Optical spectrum obtained by Fourier transform of the amplified image-encoded interferogram.

### *3.2 Imaging performance of optically-image-amplified DCM*

To evaluate the effectiveness of the optical image amplification in DCM, we first acquired the confocal amplitude image of the test chart with and without the optical image amplification. Figures 3(a) and 3(b) show the confocal amplitude image without the optical image amplification when the number of image accumulation was set to 10 and 1, corresponding to an image acquisition time of 10.2 ms and 1.02 ms or a frame rate of 97.9 fps and 979 fps, respectively. In these images, no image contrast was obtained due to too weak image-encoded signal OFC. However, when the optical image amplification was applied for DCM, the image contrast was significantly enhanced in the same acquisition condition as shown in Figs. 3(c) and 3(d). Even in the single-shot confocal amplitude imaging of Fig. 3(d), the image of the test chart was visualized moderately by the optical image amplification. In this way, comparison among them clearly indicates the effectiveness of the optical image amplification in the confocal amplitude imaging of DCM.

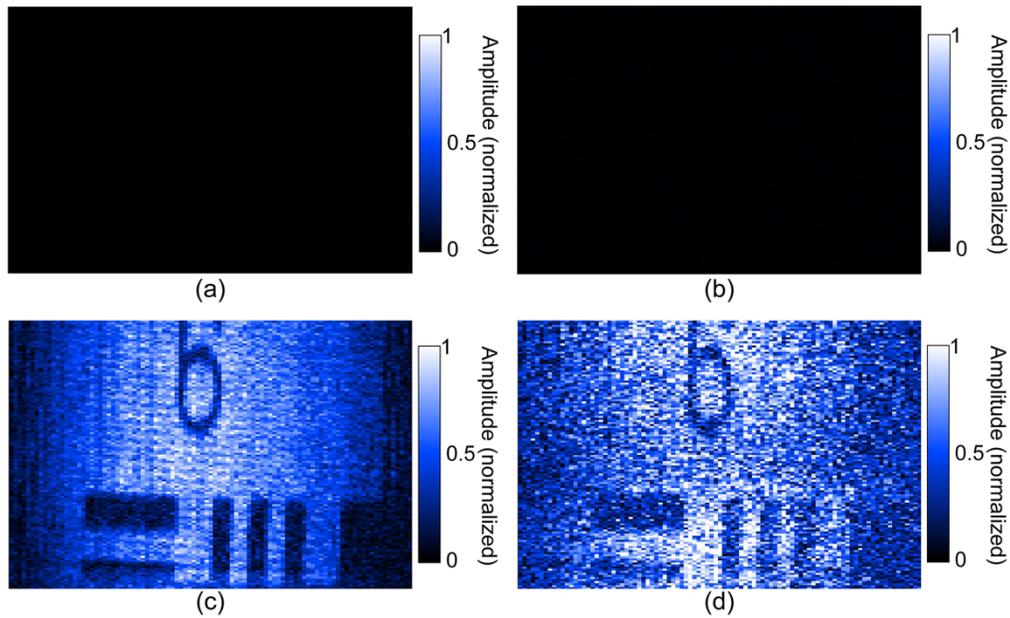

Fig. 3. Confocal amplitude image of test chart: (a) without optical amplification (number of image integration = 10, corresponding to an image acquisition time of 10.2 ms or a frame rate of 97.9 fps), (b) without optical amplification (number of image integration = 1, corresponding to an image acquisition time of 1.02 ms or a frame rate of 979 fps), (c) with optical amplification (number of image integration = 10, corresponding to an image acquisition time of 10.2 ms or a frame rate of 97.9 fps), (d) with optical amplification (number of image integration = 1, corresponding to an image acquisition time of 1.02 ms or a frame rate of 979 fps).

We next evaluate the effectiveness of the optical image amplification in confocal phase imaging of DCM. While the test chart has surface unevenness corresponding to presence or absence of reflective film, its reflectivity also depends on presence or absence of reflective film. For a reflective sample of surface unevenness with constant reflectivity, we made a thin silver coating on the test chart and used it for a sample of confocal phase imaging. Figures 4 shows comparison of the confocal phase image with and without the optical image amplification: (a) 10.2 ms or 97.9 fps and (b) 1.02 ms or 979 fps without the optical image amplification, (c) 10.2 ms or 97.9 fps and (d) 1.02 ms or 979 fps with the optical image amplification. From comparison among them, the effectiveness of the optical image amplification was confirmed again in the confocal phase imaging. It is important to note that the single-shot confocal phase imaging is also available with the help of the optical image amplification.

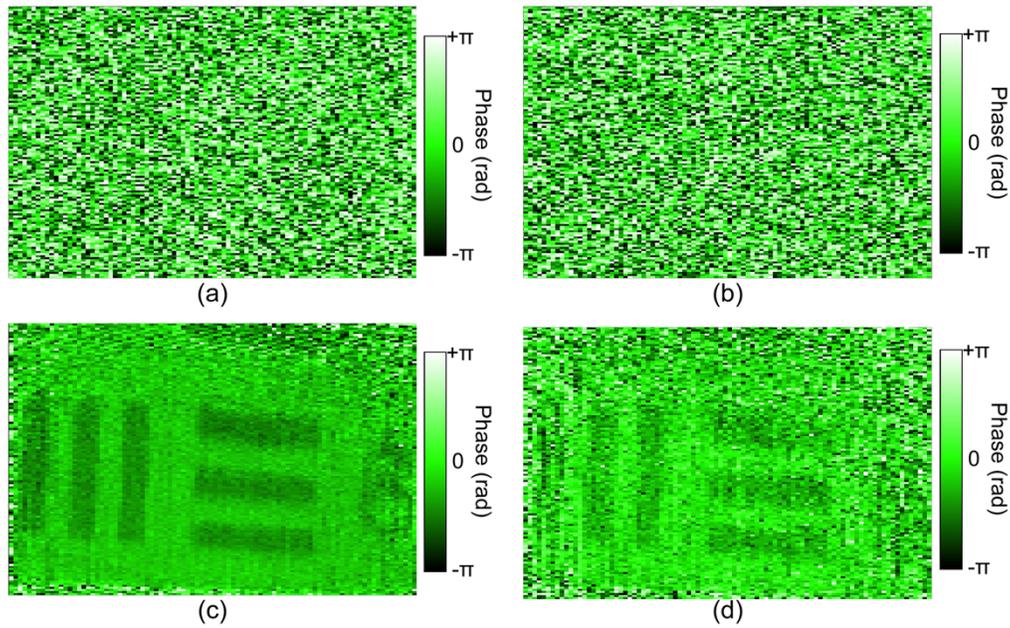

Fig. 4. Confocal phase image of test chart: (a) without optical amplification (number of image integration = 10, corresponding to an image acquisition time of 10.2 ms or a frame rate of 97.9 fps), (b) without optical amplification (number of image integration = 1, corresponding to an image acquisition time of 1.02 ms or a frame rate of 979 fps), (c) with optical amplification (number of image integration = 10, corresponding to an image acquisition time of 10.2 ms or a frame rate of 97.9 fps), (d) with optical amplification (number of image integration = 1, corresponding to an image acquisition time of 1.02 ms or a frame rate of 979 fps).

To evaluate the quantitativity of the rapid confocal phase imaging, we calculated surface unevenness $H(x, y)$ of the silver-thin-film-coated test chart by

$$H(x,y) = \frac{1}{2}\frac{\phi(x,y)}{2\pi}\lambda = \frac{\lambda}{4\pi}\phi(x,y), \tag{1}$$

where λ is a typical wavelength of OFC modes (= 1550 nm), and $\phi(x, y)$ is the phase image. We extracted cross-sectional profile of the surface unevenness from confocal phase images with different numbers of image accumulation, and determined height of the step profile with respect to number of image accumulation as shown by red plots of Fig. 5. For comparison, we measured the same sample by a white-interferometer-based 3D optical profilometer (FILMETRICS Inc., Yokohama, Japan, Profilm 3D, reflection configuration, axial resolution = 0.05 µm), and determined the height of step profile to be 63 nm as shown by a blue dash line of Fig. 5. From comparison between them, the step height was correctly determined even in the single-shot confocal phase imaging.

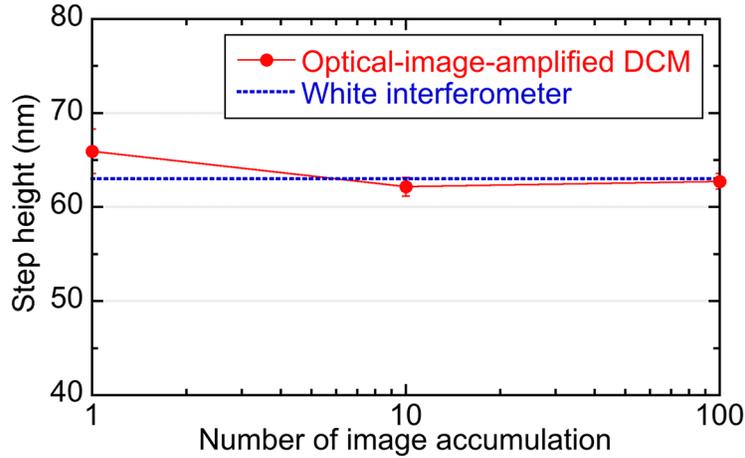

Fig. 5. Sample caption (Ref. [4], Fig. 2).

*3.3 Confocal amplitude and phase movie of moving sample*

To highlight the rapid imaging capability of the optical-image-amplified DCM, we demonstrated confocal amplitude and phase imaging of a moving sample. We prepared an aggregation of polystyrene beads with two different diameters (= 10 µm and 20 µm), and put it into water contained in a glass cell for a moving sample, as shown in Fig. 6(a). Figures 6(b) and 6(c) show snapshots of confocal amplitude and phase movie for this sample at the beginning of measurement. The corresponding movie (frame rate = 5 fps) is shown in Visualization 1. These results clearly indicated polystyrene beads were slowly swaying in all directions due to water convection rather than Brownian motion. It is important to note that Fresnel reflection from the sample surface is relatively low (= 0.75 %) because a refractive index of polystyrene beads (= 1.57 at 1550 nm) is similar to that of the water (= 1.32 at 1550 nm). Also, bright background in the confocal amplitude image and constant background in the confocal phase image are due to the Fresnel reflection from the rear surface of the glass cell because the focal point with confocal depth of 4.2 μm (not shown) was set just in front of the rear surface. Regardless of such low reflection, confocal amplitude and phase images clearly visualized temporal dynamics of polystyrene beads swaying near the rear surface of the glass cell benefiting from the optical image amplification. While the confocal amplitude image gives the reflective image of moving beads, the confocal phase image significantly gives the position of them within the range of confocal depth.

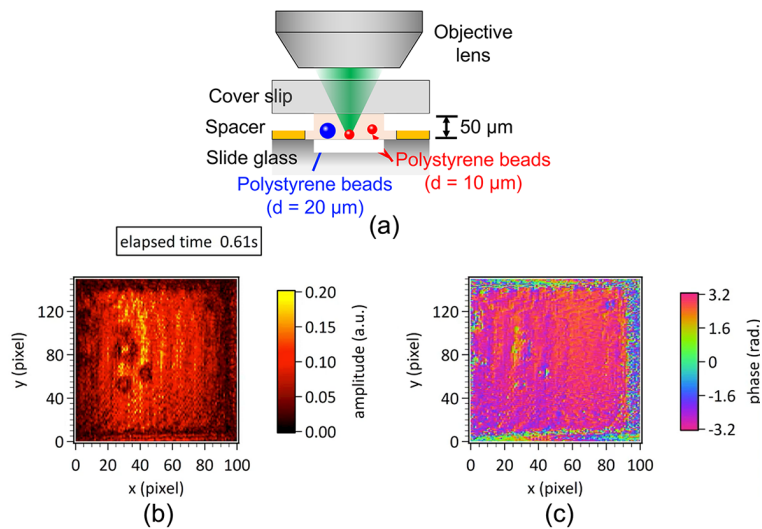

Fig. 6. (a) Schematic drawing of sample. Snapshots of confocal (b) amplitude and (c) phase movie for polystyrene beads with two different diameters slowly swaying due to water convection. The corresponding movie is Visualization 1.

## 4. Discussion

We demonstrated effectiveness of optical image amplification in DCM for enhanced quality of confocal amplitude and phase images in the rapid acquisition rate or weak signal acquisition. We here discuss comparison of optical image amplification between DCM and STEAM [24,25]. One important difference between them is in property of optical spectrum: discrete spectrum of OFC in DCM and continuous spectrum of broadband light in STEAM. Although 2D images pixels are superimposed on the optical spectrum by 2D spatial disperser and are amplified by fiber amplifiers in both DCM and STEAM, discrete optical spectrum of OFC in DCM is more robust to the pixel cross-talk in the optical image amplification than continuous spectrum in STEAM.

Another important difference between DCM and STEAM is in property of optical detection: coherent interferometric detection in DCM and incoherent non-interferometric detection in STEAM. Although the coherent interferometric detection contributes ASE rejection capability to DCM demonstrated above, incoherent non-interferometric detection makes it difficult to reject ASE because both signal light and ASE are incoherent. Also, such the coherent interferometric detection enables us to maintain the phase image contrast after optical amplification because the phase non-linearity or phase noise in optical amplification is common between the image-encoded signal OFC and the local OFC due to simultaneous process of optical amplification, leading to cancellation of the phase non-linearity or phase noise in the optical amplification process.

The final important difference between DCM and STEAM is in existence of background light. The non-interferometric detection in STEAM enables the background-free measurement and benefits from the high gain in the optical amplification if ASE is negligible. On the other hand, the background light always accompanies as non-interferometric light with the interferogram in interferometric detection in DCM, and both are optically amplified by EDFA. The amplified non-interferometric light significantly limits the dynamic range of the photodetector. As a result, the amplification ratio of EDFA was remained at 13. Such negative contribution of non-interferometric light to the photodetector dynamic range is common problem in DCS.

## 5. Conclusion

Optical image amplification was successfully introduced in DCM for confocal amplitude and phase imaging in the condition of the rapid acquisition rate or weak signal acquisition. Optical amplification of the interferogram in EDFA significantly enhances the quality of the confocal amplitude and phase images without influence of ASE background. At single-shot acquisition of confocal amplitude and phase image, high quantitative values were confirmed in nanometer-order surface topography based on the confocal phase imaging. Furthermore, temporal dynamics of polystyrene beads in water was visualized in the confocal amplitude and phase movie. The optical-image-amplified DCM will a powerful tool for real-time observation of surface topography and fast dynamic phenomena.

## Funding

Exploratory Research for Advanced Technology (ERATO), Japan Science and Technology Agency (MINOSHIMA Intelligent Optical Synthesizer Project, JPMJER1304); Japan Society for the Promotion of Science (18H01901, 18K13768, 19H00871); Cabinet Office, Government of Japan (Subsidy for Reg. Univ. and Reg. Ind. Creation); Nakatani Foundation for Advancement of Measuring Technologies in Biomedical Engineering.

## Acknowledgement

The authors acknowledge Ms. Shoko Lewis of Tokushima University, Japan, for her help in the preparation of the manuscript.**References**

1. Th. Udem, J. Reichert, R. Holzwarth, and T. W. Hänsch, "Accurate measurement of large optical frequency differences with a mode-locked laser," Opt. Lett. **24**, 881–883 (1999).

2. M. Niering, R. Holzwarth, J. Reichert, P. Pokasov, Th. Udem, M. Weitz, T. W. Hänsch, P. Lemonde, G. Santarelli, M. Abgrall, P. Laurent, C. Salomon, and A. Clairon, "Measurement of the hydrogen 1S-2S transition frequency by phase coherent comparison with a microwave cesium fountain clock," Phys. Rev. Lett. **84**, 5496–5499 (2000).

3. Th. Udem, R. Holzwarth, and T. W. Hänsch, "Optical frequency metrology," Nature **416**, 233–237 (2002).

4. S. Schiller, "Spectrometry with frequency combs," Opt. Lett. **27**, 766–768 (2002).

5. F. Keilmann, C. Gohle, and R. Holzwarth, "Time-domain mid-infrared frequency-comb spectrometer," Opt. Lett. **29**, 1542–1544 (2004).

6. T. Yasui, Y. Kabetani, E. Saneyoshi, S. Yokoyama, and T. Araki, "Terahertz frequency comb by multifrequency-heterodyning photoconductive detection for high-accuracy, high-resolution terahertz spectroscopy," Appl. Phys. Lett. **88**, 241104 (2006).

7. I. Coddington, N. Newbury, and W. Swann, "Dual-comb spectroscopy," Optica **3**, 414–426 (2016).

8. I. Coddington, W. C. Swann, and N. R. Newbury, "Coherent multiheterodyne spectroscopy using stabilized optical frequency combs," Phys. Rev. Lett. **100**, 013902 (2008).-17-